\documentstyle[12pt]{article}
\begin{document}

\vskip 1.5cm
\hskip 11.8cm {\large DTP-96/110}

\hskip 11.1cm {\large November 1996}

\vskip 2cm

\parskip=3mm
{\Large{\centerline{\bf {Testing Chiral Symmetry Breaking at DA$\Phi$NE}}}}

\vskip 1.3cm

\baselineskip=5mm
{\large{\centerline{\bf{M.R. Pennington}}}}
\vskip 5mm
{\centerline{Centre for Particle Theory,}}
{\centerline{University of Durham}}
{\centerline{Durham DH1 3LE, U.K.}}
\vskip 2.5cm
\centerline{To be published in the 
 Proceedings of the DAPHCE Workshop,}

\centerline{ held at Frascati, November 1996.}

\vskip 2cm

\centerline{A copy of the figures can be obtained from the author} 

\centerline{(m.r.pennington@durham.ac.uk).  These will be  sent by real mail.}
\newpage
.
\vspace*{2cm}

\noindent {\Large
Testing Chiral Symmetry Breaking at DA$\Phi$NE}
\vspace*{.5cm} \\
\noindent {\large  M.R. Pennington}
\vspace*{0.5cm} \\
\noindent Centre for Particle Theory, University of Durham,
Durham DH1 3LE, U.K.
\vspace*{0.4cm} \\

\baselineskip=4.8mm
\parskip=0mm
 The spontaneous breakdown of the chiral symmetry of the QCD Lagrangian
 ensures that $\pi\pi$ interactions are weak at low energies. How weak depends on the nature
 of explicit symmetry breaking.  Measurements of $K_{e4}$ decays at DA$\Phi$NE will
 provide a unique insight into this mechanism  and test whether the
 $q{\overline q}$--condensate is large or small.
 \vspace*{1cm}

\noindent{\bf  1. PIONS SCATTERING}
\vspace*{4mm}

Ever since Yukawa and certainly since Powell, pions have played a special 
role in the study of strong interactions. This is because pions are by far the
lightest of all hadrons and so, to state the obvious, determine the range 
of the nuclear force. Even before we knew why these pseudoscalar mesons should be the
lightest hadrons, the scattering of pions provided the prime process for 
studying the structure of scattering amplitudes. For instance, the high energy 
behaviour of a total cross-section is bounded with a scale set by the lightest 
exchange in the crossed channel. Since $\pi \pi $ scattering is crossing 
symmetric, the nearest $t$--channel singularity is generated by two pion exchange
and so the pion mass ( or rather $1/{m_{\pi}^{\ 2}}$ ) sets the scale for the 
Froissart bound~[1]~:
\begin{equation}
\sigma_{tot} (\pi \pi)\, \leq \,\frac{\pi}{m_{\pi}^{\ 2}}\ \ln^2(s/s_o)\quad .
\end{equation}
\noindent Now we can use the same fundamental properties of analyticity,
crossing and unitarity, to bound not just the high energy behaviour but the 
low energy too.

\begin{figure}[th]
\vspace*{11.5cm}
\caption[Fig.~1]{Mandelstam plane for $\pi\pi\to \pi\pi$. The shaded areas denote the $s$, $t$ and $u$--
channel physical regions. The symmetry point of the Mandelstam triangle is at
$s = t = u = \frac{4}{3}\ m_{\pi}^{\ 2}$.}
\end{figure}
\baselineskip=4.8mm

The natural place to start is the interior of the Mandelstam triangle. There 
though the process is unphysical, the amplitude is real (as opposed to complex)
and intimately connected to the three nearby physical regions,  Fig.~1. In the case 
of the process $\pi^0\pi^0 \rightarrow \pi^0\pi^0$, which has manifest crossing 
symmetry, the mid-point of the Mandelstam triangle at 
$s=t=u= \frac{4}{3}m_{\pi}^{\ 2}$ can be shown to be the minimum of the amplitude.
The amplitude grows in all directions~[2] away from this point because of the 
positivity of total cross-sections Fig.~2. If one is clever enough,
 and Andr\'e Martin ~[2]
is clever enough, one can deduce absolute upper and lower bounds on the
magnitude of the amplitude at this symmetry point, Figs.~1, 2. Indeed Martin~[2] found
\begin{equation}
16 > F\left(s=t=u=\frac{4}{3}m_{\pi}^{\ 2}\right) > -100\; ,
\end{equation}
\noindent with the conventional normalization of the amplitude in S-matrix 
theory. For those more familiar with applications in Chiral Perturbation Theory
($\chi$PT)~[3,4], then these values should be multiplied by $32 \pi$, i.e. $\sim$100. 
These
bounds set the size of what we would expect for a maximally strong interaction.
 The virtue of the symmetry point is that there all 
(kinematically) non-zero $\pi\pi$ amplitudes are proportional:
\begin{equation}
\frac{1}{3} F(\pi^0 \pi^0 \rightarrow \pi^0 \pi^0) = 
F(\pi^+ \pi^- \rightarrow \pi^0 \pi^0) = \frac{1}{5} F^{I=0} = 
\frac{1}{2} F^{I=2}\; .
\end{equation}

\begin{figure}[t]
\vspace*{9cm}
\baselineskip=4.80mm 
\caption[Fig.~2]{Amplitude for $\pi^0\pi^0\to \pi^0\pi^0$
over the Mandelstam triangle. The amplitude 
is a minimum at the  symmetry point, cf. Fig.~1.}
\end{figure}

\baselineskip=4.8mm
But why are pions so light? As discussed by Heiri Leutwyler here~[3] and many times
previously~[4], the {\it up} and {\it down} quarks have current masses that are very small 
compared to $\Lambda_{QCD}$. Indeed, if they were exactly massless, then the QCD 
Lagrangian would have a chiral invariance. While this is a symmetry of the 
underlying quark world, it does not occur at the hadron level --- scalars and
pseudoscalars are not degenerate for instance. Consequently, this chiral symmetry
must be spontaneously broken.

 A model that illustrates this simply is the $\sigma$--model 
 of Gell-Mann and Levy~[5] and especially that of Nambu and Jona-Lasinio~[6]. 
One considers a world of just isosinglet scalars and isotriplet pseudoscalars. 
Then the potential produced by the interactions of these fields is found to be 
a minimum not when the scalar and pseudoscalar fields are zero, but rather when 
they have a non-zero vacuum expectation value,  Fig.~3. Since there are degenerate minima, 
round the Mexican hat, nature chooses (for reasons that will become clear) a
ground state, or vacuum, near where the pion field is zero but the expectation 
value of the scalar field is definitely non-zero. The particle content is
determined by the quantum fluctuations about this vacuum. Those fluctuations in 
the pseudoscalar direction are around the rim of the Mexican hat for which there
is no resistance. Consequently, pions are massless : the Goldstone bosons of 
chiral symmetry breaking~[7]. In contrast, the fluctuations in the scalar direction
go up the sides of the Mexican hat and so the scalar field has a mass~[8]. In this
model, this comes about because the quark and antiquark fields are so strongly 
bound by non-perturbative gluon interactions that they form a condensate. The
magnitude of this condensate determines the scalar mass. It is this that gives
the (light) quarks a non-zero mass function, so that at low momenta 
$(\leq \Lambda_{QCD})$, they are effectively massive constituent quarks. In this
picture, chiral symmetry breaking not only results in massless pseudoscalars, 
but gives a non-zero mass to all the other light hadrons --- the $\rho$, the 
nucleon, etc. Their masses are determined by the scalar that acts as the Higgs 
boson of the strong interaction sector. This illustration of spontaneous chiral
symmetry breaking has close analogies with ferromagnetism, as Nambu stressed 
long ago~[9]. However, this is not the only way chiral symmetry need be broken. 
Massless Goldstone bosons always result, but a large $q {\overline q}$--condensate
is not essential. Indeed, in analogy with an antiferromagnet, the condensate 
could be small or even zero, as Jan Stern and collaborators~[10] have suggested.
It is the difference between these two pictures that DA$\Phi$NE can test.
\begin{figure}[tb]
\vspace*{11.0cm}
\baselineskip=4.80mm 
\caption[Fig.~3]{$\sigma$--model potential in the $\sigma$--${\bf \pi}$ plane. With no
explicit chiral symmetry breaking, there are degenerate minima all around the rim of the Mexican hat.
In practice, the hat is tilted so that there is a
 unique minimum close to the state labelled {\it vacuum}.}
\end{figure}

\baselineskip=4.8mm
Since pions decay by the weak interaction, they couple to an off-shell $W$--propagator 
through an axial vector current. If pions were massless, then this 
current would be conserved. This leads to an important low energy theorem. In
the massless world, when the Mandelstam triangle (Fig.~1) for $\pi\pi$ scattering 
shrinks to a point, the amplitude at this point vanishes, i.e. $F(s=t=u=0)=0$.
This obviously satisfies the bounds of Eq.~(2) (for massive pions). It highlights
how the strong interaction of pions is far weaker at low energies, than we 
would naively have expected. This allows pion amplitudes to be expanded round
the symmetry point in powers of momenta, or of the Mandelstam invariants, for
which the natural scale is the square of the mass of the scalar, or the $\rho$,
or the nucleon, or $32\pi f_{\pi}^{\ 2}$, so that
\begin{equation}
F(s,t,u,m_{\pi}=0) = \frac {{\cal O}(s,t,u)}{32\pi f_{\pi}^{\ 2}}\quad .
\end{equation}
\noindent $\chi$PT systematises this~[11,4,12].

Of course pions are not massless. There is explicit breaking of chiral 
symmetry by the masses of the current quarks and hence by the pion mass. 
Indeed, it is this explicit breaking that lifts the vacuum degeneracy of the
$\sigma$-model, so that there is a unique ground state close to the one
previously discussed. Despite this explicit breaking of chiral symmetry, 
there is still an important low energy theorem. If we consider 
$\pi\pi \rightarrow \pi\pi$ with 3 massive pions and one massless, and let the
4-momentum of the massless one go to zero, then the amplitude again vanishes.
This condition, deduced by Adler~[13], means that at the symmetry point of the 
Mandelstam triangle, Fig.~1, in the world where $s+t+u=3m_{\pi}^{\ 2}$, the amplitude
vanishes:

\begin{equation}
F\left(s=t=u=m_{\pi}^{\ 2}\right) = 0\quad .
\end{equation}

\noindent Now much closer to the real world, we can again make a Taylor 
expansion about the symmetry point in terms of not just the momenta squared,
but explicit factors of $m_{\pi}^{\ 2}$ and the value of the 
$q {\overline q}$--condensate (if this is small), all in $32\pi f_{\pi}^{\ 2}$ units. This allows 
us to consider the world with $s+t+u=4m_{\pi}^{\ 2}$. Then at the symmetry point 
we have
\begin{equation}
F\left(s=t=u= \frac{4}{3} m_{\pi}^{\ 2}\right)\, = 
\,\frac{\alpha m_{\pi}^{\ 2}}{32\pi f_{\pi}^{\ 2}}\; ,
\end{equation}
\noindent where in standard $\chi$PT with a large
 $q {\overline q}$--condensate, $\alpha$
is close to unity~[4], while in generalized $\chi$PT with a smaller condensate~[10] $\alpha$ 
can be as large as 4, when the condensate goes to zero. Thus either the 
amplitude is approximately 
\begin{equation}
F\left(s=t=u= \frac{4}{3} m_{\pi}^{\ 2}\right)\, \simeq\, 0.02 
\end{equation}
for a large condensate or 
\begin{equation}
0.04\,\leq\, F\left(s=t=u= \frac{4}{3} m_{\pi}^{\ 2}\right)\,\leq\, 0.09
\end{equation}
 for a 
smaller condensate, while the general results for a strong interaction would 
mean the amplitude is between $-100$ and $+16$, Eq.~(2). Having a small value for
the $\pi\pi$ amplitude at the symmetry point speaks of the Goldstone nature of
the pion. How small this value is tells us about the explicit breaking of
chiral symmetry. It is at the symmetry point that the difference is maximal. 
This is because moving closer to the physical regions the parameters of the
Chiral Lagrangians $({\overline \ell_i}, \lambda_i)$ [4,10] are necessarily fixed from experiment
and the different expansions inevitably become more similar. Thus it is the
amplitude at the symmetry point, Eqs.~(7,8), that distinguishes the versions of explicit 
breaking.

To learn about this, we must continue experimental information into the
unphysical region, Fig.~1. The way to do this is by using dispersion relations. The 
$\pi\pi$ amplitudes satisfy fixed-$t$ dispersion relations that, thanks to the 
Froissart bound, need at most two subtractions. This would mean there would 
be 2 subtraction constants at each momentum transfer. However, by the clever
use of crossing symmetry S.M. Roy~[14] showed how these could be rewritten in terms
of just two subtraction constants for a range of momentum transfers --- a range 
that allows a rigorous type of partial wave dispersion relations, known as Roy
equations. It is natural to take the two subtraction constants to be values of the 
$I=0$ and $I=2$ $S$--wave amplitudes at threshold, which give the scattering lengths 
$a_{0}^{0}$ and $a_{0}^{2}$. While, in general, these are independent, in the
real world these two constants are closely correlated. This is because all 
experiments and models have an $I=1$ cross-section below 1 GeV dominated by the
$\rho$-resonance and have an $I=2$ cross-section that is comparatively small~[15].
This forces $a_{0}^{0}$ and $a_{0}^{2}$ to follow what is known as the universal 
curve of Morgan and Shaw~[16], along which all phase-shift solutions and all models 
lie. Standard $\chi$PT~[17] gives $a_{0}^{0}=0.21 \pm 0.01$, for example, at 2 loops 
with electromagnetic corrections, while in generalized $\chi$PT $a_{0}^{0}$ is
bigger~[18]. For instance if $\alpha = 3$ (Eq.~(6)), then $a_{0}^{0} \simeq 0.31$.
Though the one parameter $\pi\pi$ scattering depends on can be conveniently taken
to be the $I=0$ $S$--wave scattering length, $a_{0}^{0}$, it could equally well be
the value of the $\pi\pi$ amplitudes at the symmetry point.
As just noted changing this value by a factor 3, $a_{0}^{0}$ only 
changes by 50\%. What do data on $\pi\pi$ scattering, tell us about this 
scattering length?


\begin{figure}[p]
\vspace*{16.5cm}
\baselineskip=4.80mm 
\caption[Fig.~4]{$S$--wave $\pi\pi$ phase-shifts $\delta_0^{\ I}$ with isospin $I=0, 2$
as a function of $\pi\pi$ mass.  The data are from
different analyses of the CERN-Munich experiments~[19,20].
For $\delta^0_0$, the open triangles below 620 MeV are from 
Estabrooks and Martin~[21] (averaging their $s$ and $t$--channel treatments).  Above 610 MeV are shown results of the
energy-independent (open circles) and energy-dependent
 (solid triangles) analyses of Ochs~[22].
The $I=2$ phases, $\delta^2_0$, in 100 MeV bins are the results of the
two analyses of Hoogland et al.~[20]~: method A (open circles),
method B (solid squares). The solid line marks what are called the {\it central phases} and their extrapolation to threshold using the results of the study
of the Roy equations with $a^0_0 =0.2$, the dotted line has $a^0_0 =0.1$ and the dashed line
$a^0_0 =0.3$.}
\end{figure}


\baselineskip=4.8mm
\vspace*{5mm}

\noindent {\bf 2. PION INTERACTIONS FROM EXPERIMENT}
\vspace*{5mm}

The classic CERN-Munich experiments~[19,20] on dipion production at high energies and 
small momentum transfers for $\pi^{\pm}p \rightarrow \pi^{\pm}\pi^{+}n$ give the 
$\pi\pi$ phase-shifts~[20-22] for the $I=0$ and 2 $S$--waves, Fig.~4. While these data
 clearly
indicate scattering lengths smaller than 1 (in pion mass units), 
as seen from the Roy equation solutions in Fig.~4 they allow a
range of values for $a_{0}^{0}$ far larger than the tiny difference
(corresponding to Eqs.~(7,8)) we want to be able to
detect if we are to test the nature of explicit chiral symmetry breaking. 
It is clear one needs 
precision data closer to threshold than 500 MeV. These will be provided by 
$K_{e4}$ decays at DA$\Phi$NE~[23]. With a branching ratio of 
$4.10^{-5}$, $\ K^{\pm} \rightarrow e^{\pm}\nu\ \pi^+ \pi^-$. Here, the pions 
interact in the femto-universe of the decay in a universal way and they remember
how they interact. This means that each $\pi\pi$ partial wave amplitude has the 
same phase as it does in all $\pi\pi$ interactions, and so as in elastic
scattering. It is these phases, or rather their differences, we want
to measure. This is not an easy task, since the process of a decay into 4 
particles, e.g. $K^+ \rightarrow e^+ \nu_e\ \pi^+ \pi^-$, depends on 5 Lorentz 
invariants. These can be conveniently expressed in terms of 5 experimentally 
measurable kinematic variables. As proposed by Cabibbo and Maksymowicz~[24], we 
imagine the $K$--decays into a dilepton and a dipion system back-to-back
shown in Fig.~5 and the
masses of these two systems $M_{e\nu}$ and $M_{\pi\pi}$ are the first two
variables. Next we consider the dilepton rest frame, where the individual 
leptons go off back-to-back at an angle $\theta_e$ relative to the initial
dilepton direction. Similarly, the pions go off at an angle $\theta_{\pi}$
relative to the initial dipion system. Then lastly, the plane of the two leptons
is at an angle $\phi$ to the two pion plane, Fig.~5. The decay distribution is then 
studied as a function of $M_{e\nu} , M_{\pi\pi} , \theta_e , \theta_{\pi}$
and $\phi$. Results from DA$\Phi$NE will not be limited by statistics alone, but
as much or more by systematics. These can be ameliorated by having large samples of 
both $K^+$ and $K^-$'s. These decays are related by a $CP$ transformation, so
that their distributions should be the same, except that the dilepton and dipion
planes are oppositely oriented, i.e. $\phi \rightarrow -\phi$. This fact can
help check how well the $\phi$--dependence of the KLOE detector is understood and 
so reduce systematic uncertainties~[25].

\begin{figure}[th]
\vspace*{11.5cm}
\baselineskip=4.80mm
\caption[Fig.~5]{$K^+\to e^+ \nu_e\ \pi^+\pi^-$ decay : (a) showing the weak decay of the strange quark,
(b)~the kinematics discussed in the text.}
\end{figure}

\begin{figure}[th]
\vspace*{11.3cm}
\baselineskip=4.80mm
\caption[Fig.~6]{The phase difference $\delta^0_0 - \delta^1_1$ determined from
$K_{e4}$ decays in the Pennsylvania (open triangles)~[29] and Geneva-Saclay
(solid dots)~[30]
experiments as a function of $\pi\pi$ mass.  The curves are the predictions of 
two loop $\chi$PT~:
the Standard result~[17] for which $\alpha \simeq 1.2$ and for $\alpha=2, 3$ of Generalised
$\chi$PT~[18].}
\end{figure}

\baselineskip=4.8mm

\baselineskip=4.8mm
Now the way the decays depend on the kinematic variables is specified by the 
weak matrix element for $K^{\pm} \rightarrow \pi^+ \pi^- \ W^{\pm} \rightarrow 
\pi^+ \pi^- e^{\pm}\nu$. The $W^{\pm}$ are described by vector and axial vector 
currents. Their matrix elements can be expressed in terms 
of the Lorentz vectors~[24]
that can be formed from the kaon and the two pion momenta. Each independent
combination is multiplied by a formfactor $F$, $G$, $R$ and $H$, which will depend on
$M_{\pi\pi}$ and $M_{e\nu}$. Taking the modulus squared of these matrix elements 
and using the Dirac equation, the formfactor $R$ is multiplied by 
$m_e^2 / M_{e\nu}^2$, so if we avoid the difficult measurements at very small dilepton 
masses, this term is negligible. Consequently, the 5-fold differential decay
distribution depends on the unknown formfactors $F$, $G$ and $H$. These can be 
decomposed in terms of $\pi\pi$ partial waves. Analysis shows that $D$ and higher 
waves are negligible~[26], so that only $S$ and $P$--waves are needed. The decay 
distribution involves terms that depend on the $S-P$ phase difference, 
$\delta_S - \delta_P$. By Watson's final state interaction theorem and the 
$\Delta I = \frac{1}{2}$ rule, this difference is just equal to the difference
between the $I=0$ $S$--wave and $I=1$ $P$--wave phases, $\delta_0^0 - \delta_1^1$, of
$\pi\pi$ elastic scattering. Since $\delta_1^1$ is accurately determined from 
the tail of the $\rho$ by the use of dispersion relations, these measurements
allow $\delta_0^0$ to be determined. As seen from Eq.~(20) of [27] and Eq.~(5.19) of
[28], it is the 
$\phi$--dependence that mainly determines this phase difference. 
Pais and Treiman~[27]
proposed a method to determine this independently
of the unknown formfactors.  The method involves the double differential
decay distribution for $\cos \theta_e$ and $\phi$. However the crucial
$\phi$--dependence is experimentally very weak and the use of information
on just two variables is insufficient for a precision determination.
Rather a maximum 
likelihood analysis of the full five-fold differential decay
distribution is needed. With 7000 events from the University of
 Pennsylvania group~[29]
 and 30000 events from the Geneva-Saclay group~[30] give the phase differences shown
in Fig.~6.   A free fit to the Geneva-Saclay  results yields 
$a_0^0 = 0.31 \pm 0.11$, while incorporating the Roy 
equation constraints gives~[15]
\begin{equation}
a_0^0 = 0.28 \pm 0.05 \hspace{1cm}{\rm or}\hspace{1cm}  a_0^0 = 0.26 \pm 0.05
\end{equation}
\noindent depending on the solutions used. These can be translated into  values
of the amplitude at the symmetry point, Eq.~(6), with~[10,18]
\begin{equation}
1.30 \leq \alpha \leq 3.02
\end{equation}
\noindent for the  second values of $a_0^0$ in Eq.~(9). Clearly, this range cannot 
distinguish between the different forms of explicit chiral symmetry breaking
and tell whether the $q {\overline q}$--condensate is large or small. The aim is that
with higher statistics and better control of systematics, DA$\Phi$NE will do
better. Clearly greater precision on the phases would be achieved if the 
formfactors $F$, $G$, $H$ were modelled as in $\chi$PT~[28], but this would be prejudicing 
the result. Further study will be needed in the specific context of the KLOE
detector~[23,25].

\begin{figure}[th]
\vspace*{8.2cm}
\baselineskip=4.80mm 
\caption[Fig.~7]{Unnatural parity exchange in $\pi N\to\pi\pi N$
 at high energy and small momentum transfer.}
\end{figure}

\baselineskip=4.8mm
If the $q {\overline q}$--condensate  is large, then in the NJL model~[6], the scalar 
mesons form the Higgs sector of the strong interaction. Much of our present
information about this comes from the 25 year old CERN-Munich experiments~[19,20],
mentioned earlier. These were first analysed assuming the 
unnatural parity exchange component  of the $\pi$N $\rightarrow \pi\pi$N cross-section at 
17.2 GeV/c
to be controlled by $\pi$ exchange alone, Fig.~7. This gives the Ochs and Wagner~[22], or 
Estabrooks and Martin~[21] phases shown in Fig.~4. A test of these assumptions was provided by the
later ACCMOR collaboration measurements~[31] on a polarized target. These showed
that the unnatural parity component
 was consistent with one-pion-exchange
below 1 GeV. However, taking new data at 5.95 and 11.85 GeV/c~[32] and using the ACCMOR results 
up to 900 MeV, Svec~[33] claimed that an {\it up} solution~[15] for $\delta_0^0$ was also
possible, indeed favoured, indicating a narrow \lq\lq$\sigma$" with $\rho$-like mass 
and width. It was known that this could not possibly be correct~[15], since this
solution could not describe the sharp change in behaviour of the integrated
cross-section and $S-P$ interference at $K {\overline K}$ threshold. These require the 
phase $\delta_0^0$ to rise steeply through $180^\circ$ near $K {\overline K}$ threshold
and the phase could not have already reached this value below 900 MeV as Svec 
proposed. Nevertheless, these claims prompted the Krakow group~[34] to go back and
reanalyse their data from ACCMOR taken nearly twenty years ago. They find one solution
(the so called {\it down-flat} solution, Fig.~8a) which is very close to that of Ochs and 
Wagner~[22], and of Estabrooks and Martin~[21] of Fig.~4. However, they do find a second solution 
({\it up-flat}, Fig.~8b) that does give a steeper rise of the phase below 900 MeV and a not so
steep rise through $K {\overline K}$ threshold. (There are incidentally 2 other 
solutions but they are not consistent with unitarity.) These two solutions must
have different $\pi$ and other unnatural parity exchange, e.g. $a_1$, 
components, Fig.~7. 
\begin{figure}[th]
\vspace*{10cm}
\baselineskip=4.80mm 
\caption[Fig.~8] {The {\it down-flat} and {\it up-flat} solutions for
the $I=0$ $S$--wave phase-shift, $\delta_0^0$, as functions of $\pi\pi$ mass,
from the recent re-analysis  of the CERN-Munich and ACCMOR data [19,31].
The {\it down-flat} solution is very similar to that displayed in Fig.~4 from
Refs.~[21,22].}
\end{figure}

These can be tested by measuring 
$\pi^- p \rightarrow \pi^0 \pi^0n$ on a polarized target and extracting the
$\pi^- \pi^+ \rightarrow \pi^0 \pi^0$ amplitude.
The advantage of this
$\pi^0\pi^0$ channel 
is that there are only even $\pi\pi$ partial waves, with no $\rho$-contribution, 
but no data on a polarized target are likely to be taken. Nonetheless, the 
Krakow group can predict for their solutions what the $S$--wave cross-sections
for unnatural parity exchange in an unpolarized $\pi^-p \rightarrow \pi^0\pi^0n$ 
experiment at 17.2~GeV/c should be. Fortunately, the E852 collaboration at BNL~[35] has
measured this at 18~GeV/c already, as Alex Dzierba~[36] has described here. So if the 
Krakow group make these predictions, E852, once their full statistics are 
analysed, should be able to distinguish between these solutions. This is an essential
test of the $S$--wave $\pi\pi$ amplitudes, on which any view of the scalar mesons
is necessarily based.
Both DA$\Phi$NE~[23] at Frascati and CEBAF~[36] at TJNAF can add to this by their 
radiative $\phi$-decay experiments. With sufficient statistics the $\pi\pi$
mass spectrum can be mapped out and this too will aid our understanding of the
universality of $\pi\pi$ interactions.

The explicit breaking of chiral symmetry can only be tested in very low energy $\pi\pi$
processes. Measurements of $K_{e4}$ decays at DA$\Phi$NE~[23], or of the 
lifetime of $\pi^+\pi^-$ atoms~[37] at CERN, alone can achieve this.  These have the potential to provide a unique
insight into one of the most fundamental properties of QCD.

\vspace*{0.5cm} 
\noindent {\Large{Acknowledgements}}
\vspace*{0.2cm}

Participation at this meeting was made possible by the 
 support of the EC Human Capital and Mobility Programme which funds the
{\it EURODA$\Phi$NE} network under grant CHRX-CT920026.

\newpage

\end{document}